\documentclass[12pt]{article}
\usepackage[square]{natbib} 
\newcommand{\keywords}[1]{\textbf{Keywords:} #1}
\usepackage[ruled,vlined]{algorithm2e}
\usepackage{color}
\usepackage{subcaption}
\usepackage{amsthm}
\usepackage{hyperref}
\usepackage{graphicx}
\usepackage{amsfonts}
\usepackage{amsmath}
\usepackage{wasysym}
\usepackage{url}
\usepackage{amssymb}
\usepackage{multirow}
\usepackage{tabularx} 
\usepackage{bm}
\usepackage{booktabs}
\usepackage{url}
\usepackage{mathrsfs}
\usepackage{fullpage}
\usepackage{setspace}
\usepackage[english]{babel}
\usepackage[utf8]{inputenc}
\usepackage{bibunits}

\newtheorem{theorem}{Theorem}
\newtheorem{proposition}{Proposition}

\newtheorem{definition}{Definition}
\def\bbE{\mathbb{E}}

\def\bbR{\mathbb{R}}

\usepackage{xr}
\title{\textbf{Community detection in heterogeneous signed networks}}

\author{
	Yuwen Wang$^\dag$, Shiwen Ye$^\dag$, Jingnan Zhang$^\ddag$, Junhui Wang$^\dag$ \\ [10pt]
	$^\dag$Department of Statistics and Data Science \\
	The Chinese University of Hong Kong \\
        $^\ddag$Faculty of Business for Science \& Technology\\
        School of Management\\
	University of Science and Technology of China\\
}
\date{}

\begin{document}

\maketitle
\onehalfspacing
\begin{abstract}
Network data has attracted growing interest across scientific domains, prompting the development of various network models. Existing network analysis methods mainly focus on unsigned networks, whereas signed networks, consisting of both positive and negative edges, have been frequently encountered in practice but much less investigated. In this paper, we formally define strong and weak balance in signed networks, and propose a signed block $\beta$-model, which is capable of modeling strong- and weak-balanced signed networks simultaneously. We establish the identifiability of the proposed model by leveraging properties of bipartite graphs, and develop an efficient two-step algorithm to tackle the resulting optimization problem and recover the community membership. More importantly, we establish the asymptotic consistencies of the proposed model in terms of both parameter estimation and community detection. Its advantages are also demonstrated through extensive numerical experiments and the application to a real-world international relationship network. 
\end{abstract}

\medskip

\newpage

\doublespacing

\section{Introduction}

Network data has attracted increasing attention in recent years, with applications in diverse scientific domains, ranging from social science \citep{hunter2008goodness}, computer science \citep{heard2014filtering} to biomedicine \citep{shojaie2009analysis}. Analysis of network data can reveal latent structures underlying the interactions among entities of interest, such as community detection \citep{girvan2002community}, link prediction \citep{martinez2016survey}, homophily \citep{hoff2007modeling} and so on. 

\begin{figure}[!htp]
    \centering
    \includegraphics[width=0.5\linewidth]{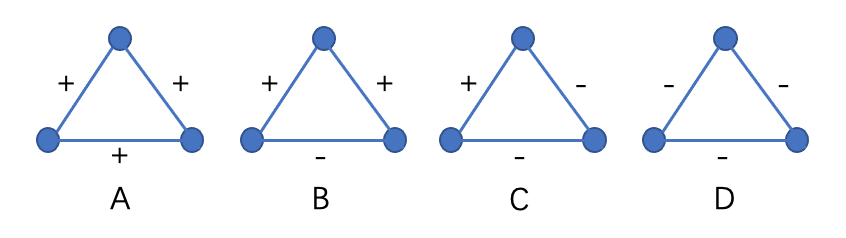}
    \caption{Configurations of four possible triads.}
    \label{fig:triads}
\end{figure}

Most existing methods for network analysis focus on unsigned networks, whereas signed networks are also prevalent in real-world applications. Specifically, signed networks have been extensively employed to represent relationships among objects \citep{harary1953notion}, which associate polarity information with each edge and thus distinguish friends from enemies, or allies from rivals. These signed edges give rise to unique structural characteristics grounded in social psychology theories, particularly the balance theory \citep{heider1946attitudes, cartwright1956structural}, stating that certain triadic configurations involving signed edges lead to stable or unstable social structures. Figure \ref{fig:triads} illustrates four possible triads, among which triads A and C represent strong-balanced structures. These triads align with intuitive social principles, such as ``a friend of my friend is likely my friend'' and ``an enemy of my friend is likely my enemy''. Weak balance \citep{davis1967clustering} extends this concept by additionally permitting triad D, which accommodates the occurrence of three mutually hostile nodes and is thus more suitable for modeling real-life signed networks. 

Community detection has been widely studied for unsigned networks \citep{holland1983stochastic, girvan2002community, newman2006modularity, lei2015consistency,zhang2022directed}, but much less so for signed networks. Most existing community detection methods for signed networks \citep{kunegis2010spectral,jiang2015stochastic,cucuringu2021regularized} are algorithm based and lack formal theoretical analysis, not to mention that the prevalent balanced structures in signed networks are often ignored by the aforementioned community detection methods. Note that the balance theory actually posits that a strong-balanced (weak-balanced) network can be partitioned into two (multiple) sub-networks such that positive edges connect nodes within the same sub-network, whereas nodes from different sub-networks are connected by negative edges \citep{easley2010networks}. More recently, \cite{tang2024population} introduces a formal definition of strong-balanced signed networks in a probabilistic sense and proposes an efficient method to detect the two communities, yet the method is inapplicable to analyze the more realistic weak-balanced signed networks with multiple communities. 

In this paper, we first provide a new set of conditions for both strong- and weak-balanced signed networks, from both local and global perspectives. Built upon the global perspective, we then propose a signed block $\beta$-model to explicitly incorporate the strong or weak balance structures in heterogeneous signed networks. Particularly, it introduces two sets of parameters to capture each node's distinct connection tendencies within its own community and across different communities. The identifiability of the signed block $\beta$-model is established by exploiting properties of the bipartite graph, and an efficient two-step algorithm is developed to solve the resulting optimization problem and recover the community membership. More importantly, the asymptotic consistencies of the signed block $\beta$-model are established in terms of both parameter estimation and community detection. Its advantages are also supported in numerical experiments on various synthetic signed networks and a real-world international relationship network.

\paragraph{Notations.} We introduce some notations to be used throughout the paper. For any integer $n$, let $[n]=\{1,\dots,n\}$. For any vector $\bm{v}=(v_1,\dots,v_n)^{\top} \in\mathcal{R}^n$, $\|\bm{v}\|_2^2=\sum_{i\in[n]}v_i^2$, and $\|\bm{v}\|_{\infty}=\max_{i\in[n]}|v_i|$. We use $\bm{1}_n$ to denote the $n$-dimensional column vector with all elements being 1 and $\mathbf{1}(\cdot)$ to denote the indicator function. For any matrix $\bm{A} = (a_{ij}) \in\mathcal{R}^{n\times m}$, let $\bm{A}_{i\cdot}$ be its $i$-th row, $\bm{A}_{\cdot i}$ be its $i$-th column, and $\|\bm{A}\|_F^2=\sum_{i\in[n],j\in[m]}a_{ij}^2$ be its Frobenius norm. For any two matrices $\bm{A}$ and $\bm{B}$, $\bm{A}\odot\bm{B}$ is the Hadamard product with $(\bm{A}\odot\bm{B})_{ij}=a_{ij}b_{ij}$. Let $\bm J = I_n-\frac{\bm 1_n\bm 1_n^\top}{n}$ be the centering matrix. For any two vectors $\bm x=(x_1, \dots, x_n)$ and $\bm y=(y_1, \dots, y_n)$, $\bm x<\bm y$ denotes $x_i<y_i$ for each $i\in[n]$. For any two sequences $\{a_n\}_{n\ge 1}$ and $\{b_n\}_{n\ge 1}$, we denote $a_n \prec b_n$ if $a_n=o(b_n)$, $a_n \preceq b_n$ if $a_n=O(b_n)$ and $a_n\simeq b_n$ if $a_n \preceq b_n$ and $b_n \preceq a_n$. For any set $\mathcal{X}$, $|\mathcal{X}|$ denotes its cardinality and $\mathcal{X}^c$ denotes its complement. For any two sets $\mathcal{X}$ and $\mathcal{Y}$, $\mathcal{X} \setminus \mathcal{Y} = \{ x \mid x \in \mathcal{X} \text{ and } x \notin \mathcal{Y} \}$.

\section{Signed network and balance structure}

A signed network $\mathcal{G}$ can be represented via its symmetric adjacency matrix $\bm{A} = (a_{ij})_{i,j\in[n]}$ with $a_{ij} \in \{1, -1, 0\}$, where $a_{ij} = 1$ indicates a positive edge between nodes $i$ and $j$, $a_{ij} = -1$ indicates a negative edge and $a_{ij}=0$ indicates no edge at all. For each $i, j\in[n]$, let $a_{ij}$ be sampled from a multinomial distribution with parameters $(p_{ij}^+, p_{ij}^-, p_{ij}^{0})$. One of the most salient features of signed network is the widely observed balance structure \citep{harary1953notion, davis1967clustering}, positing that relationships among entities tend toward balanced sign configurations. 

\begin{definition} [Strong balance]
\label{def:strong balance}
    A signed network is strong-balanced if for any triple $(i, j, k)$, 
    $$
    \underset{(a, b, c)\in\{-1, 1\}^3}{\arg\max} \mathbb{P} \big (a_{ij}=a, a_{jk}=b, a_{ki}=c \big )
    $$
    does not have 0 or 2 positive edges.
\end{definition}

Definition \ref{def:strong balance} assures that a strong-balanced signed network does not contain a triad achieving the highest probability at triad B or D as shown in Figure \ref{fig:triads}. It can also be extended to define weak-balanced signed networks.


\begin{definition}[Weak balance]
\label{def:weak balance}
    A signed network is weak-balanced if for any triple $(i, j, k)$, 
    $$
    \underset{(a, b, c)\in\{-1, 1\}^3}{\arg\max} \mathbb{P} \big (a_{ij}=a, a_{jk}=b, a_{ki}=c \big )
    $$
    does not have exactly 2 positive edges.
\end{definition}

Definition \ref{def:weak balance} gives a probabilistic definition of weak balance for signed networks. It only restricts the occurrence of triad B, and thus aligns with the non-probabilistic definitions of weak balance in the existing literature \citep{easley2010networks}.  

\begin{proposition} \label{prop:equ-def-sbalance}
A signed network $\mathcal{G}$ is strong-balanced if and only if

(i) For any $(i, j, k)$, $(p_{ij}^+-p_{ij}^-)(p_{jk}^+-p_{jk}^-)(p_{ki}^+-p_{ki}^-)>0$; or

(ii) There exists a partition \( [n] = C_1 \cup C_2 \) such that for each $i,j\in[n]$, $p_{ij}^+ > p_{ij}^- $ if $i,j$ belong to the same subset, and $p_{ij}^+ < p_{ij}^-$, if  $i,j$  belong to different subsets.
\end{proposition}

\begin{proposition} \label{prop:equ-def-wbalance}
A signed network $\mathcal{G}$ is weak-balanced if and only if

(i) For any $(i, j, k)$, $(p_{ij}^+-p_{ij}^-)(p_{jk}^+-p_{jk}^-)(p_{ki}^+-p_{ki}^-)>0$ or $p_{ij}^+-p_{ij}^-<0, p_{jk}^+-p_{jk}^-<0, p_{ki}^+-p_{ki}^-<0$; or

(ii) There exists a partition of $[n]=C_1\cup C_2\cup ...\cup C_K$ with $K\ge 3$ such that for each $i,j\in[n]$, $p_{ij}^+ > p_{ij}^- $ if $i,j$ belong to the same subset, and $p_{ij}^+ < p_{ij}^-$, if  $i,j$  belong to different subsets.
\end{proposition}

Propositions \ref{prop:equ-def-sbalance} and \ref{prop:equ-def-wbalance} characterize the strong and weak balance structures in signed networks from both local and global perspectives. In fact, the local perspective (i) in Proposition \ref{prop:equ-def-sbalance} is essentially equivalent to $\bbE (a_{ij}a_{jk}a_{ki}\big| |a_{ij}a_{jk}a_{ki}| = 1)>0$ for any $(i, j, k)$, which is exactly the definition of the strong-balanced signed network in \cite{tang2024population}, whereas Proposition \ref{prop:equ-def-wbalance}(i) can be regarded as its analogy for the weak-balanced signed network.  On the other hand, Proposition \ref{prop:equ-def-sbalance}(ii) and \ref{prop:equ-def-wbalance}(ii) give the brand new global perspectives on strong and weak balance structures based on partitions of the signed network, which greatly facilitate the modeling of community structures in signed networks.

\section{Signed block $\beta$-model}

We propose a signed block $\beta$-model, denoted as SBBM, to leverage the global perspectives of the balance structures in signed networks. Particularly, let $\sigma:[n]\rightarrow[K]$ denote the community membership of node $i$, then 
\begin{equation}
\label{Model: sgbeta}
\begin{aligned}
    \mathbb{P}(a_{ij}=1)&=\begin{cases}
        \frac{e^{ \beta_i^++\beta_j^+}}{e^{ \beta_i^++\beta_j^+}+e^{ \beta_i^-+\beta_j^-}+1},\ \text{if\ } \sigma(i)=\sigma(j)\\
        \frac{e^{ \gamma_i^++\gamma_j^+}}{e^{ \gamma_i^++\gamma_j^+}+e^{ \gamma_i^-+\gamma_j^-}+1},\ \text{if\ } \sigma(i)\neq \sigma(j)
        \end{cases},\\
    \mathbb{P}(a_{ij}=-1)&=\begin{cases}
        \frac{e^{ \beta_i^-+\beta_j^-}}{e^{ \beta_i^++\beta_j^+}+e^{ \beta_i^-+\beta_j^-}+1},\ \text{if\ } \sigma(i)=\sigma(j)\\
        \frac{e^{ \gamma_i^-+\gamma_j^-}}{e^{ \gamma_i^++\gamma_j^+}+e^{ \gamma_i^-+\gamma_j^-}+1},\ \text{if\ } \sigma(i)\neq \sigma(j)
    \end{cases},
\end{aligned}
\end{equation}
for each $1\le i\le j\le n$, and $\mathbb{P}(a_{ij}=0)=1-\mathbb{P}(a_{ij}=1)-\mathbb{P}(a_{ij}=-1)$. It is clear that SBBM associates each node $i$ with two sets of parameters $(\beta_i^+, \beta_i^-)$ and $(\gamma_i^+,\gamma_i^-)$, where $(\beta_i^+,\beta_i^-)$ quantifies the node's tendency to form positive and negative edges within the same community, and $(\gamma_i^+,\gamma_i^-)$ characterizes the corresponding tendencies across different communities. 

The proposed SBBM model has a number of compelling features. First, it integrates the capability of community detection while simultaneously capturing node-specific heterogeneity in signed networks. When $K=1$ and $\beta_i^-=-\infty$, SBBM reduces to the classical $\beta$-model, which is commonly used to capture node-specific heterogeneity. In the absence of negative edges, SBBM simplifies to the stochastic block model (SBM) with heterogeneity for community detection. Therefore, SBBM provides a general framework for both unsigned and signed networks, simultaneously addressing community detection, node-specific heterogeneity and signed relationships in network data. Second, both strong and weak balance can be incorporated in SBBM by simply requiring
\begin{equation}
\label{Model:sign inq}
    \beta_i^+>\beta_i^- \mbox{ and } \gamma_i^+<\gamma_i^-\ \text{ for any } i\in [n].
\end{equation}
It can be readily verified that nodes $i$ and $j$ belong to the same community if and only if $\mathbb{P}(a_{ij}=1)>\mathbb{P}(a_{ij}=-1)$, and different communities if and only if $\mathbb{P}(a_{ij}=1)<\mathbb{P}(a_{ij}=-1)$. It follows from Propositions \ref{prop:equ-def-sbalance} and \ref{prop:equ-def-wbalance} that SBBM with constraints \eqref{Model:sign inq} satisfies strong balance when $K=2$ and weak balance when $K\geq 3$. In contrast, enforcement of strong and weak balance structures can be challenging in most other existing signed network models.

\subsection{Model identifiability}

Let $\bm{Z}=(\bm{z}_1,\dots,\bm{z}_n)^\top$ be the community membership matrix, where $\bm{z}_i\in\{0,1\}^K$ with $z_{ik}=1$ if $\sigma(i)=k$. Define $\bm{\beta}^+ = (\beta^+_{i})_{i=1}^n$, $\bm{\gamma}^+=(\gamma_i^+)_{i=1}^n$, $\bm{\beta}^- = (\beta^-_{i})_{i=1}^n$, $\bm{\gamma^-}=(\gamma_i^-)_{i=1}^n$, $\bm\eta^+=\bm\beta^+-\bm\gamma^+$ and $\bm\eta^-=\bm\beta^--\bm\gamma^-$. Then, the parameter space is
\begin{align*}
	\mathcal{P}=\big\{(\bm\gamma^+,\bm\eta^+,\bm\gamma^-,\bm\eta^-,\bm{Z})\in\mathbb{R}^n\times\mathbb{R}^n\times\mathbb{R}^n&\times\mathbb{R}^n\times\{0,1\}^{n\times K}:\\ &\eqref{Model:sign inq}\ \text{is satisfied and}\ \bm{Z}\bm{1}_K=\bm{1}_n \big\},
\end{align*}
where $\bm{Z}\bm{1}_K=\bm{1}_n$ assures that each node belongs to exactly one community. Moreover, let $\bm\Theta^+=(\theta_{ij}^+)_{i,j\in[n]}$ and $\bm\Theta^-=(\theta_{ij}^-)_{i,j\in[n]}$, where $\theta_{ij}^+=\log\frac{\mathbb{P}(a_{ij}=1)}{1-{\mathbb{P}(a_{ij}=1)-\mathbb{P}(a_{ij}=-1)}}$ and $\theta_{ij}^-=\log\frac{\mathbb{P}(a_{ij}=-1)}{1-\mathbb{P}(a_{ij}=1)-{\mathbb{P}(a_{ij}=-1)}}$. Then, SBBM in \eqref{Model: sgbeta} can be rewritten as 
\begin{align*}
\begin{cases}
\bm\Theta^+=\bm\gamma^+\bm{1}_n^\top+\bm{1}_n\bm\gamma^{+\top}+(\bm\eta^+\bm{1}_n^\top+\bm{1}_n\bm\eta^{+\top})\odot\bm{Z}\bm{Z}^\top,\\
\bm\Theta^-=\bm\gamma^-\bm{1}_n^\top+\bm{1}_n\bm\gamma^{-\top}+(\bm\eta^-\bm{1}_n^\top+\bm{1}_n\bm\eta^{-\top})\odot\bm{Z}\bm{Z}^\top.
\end{cases}
\end{align*}

We say SBBM is identifiable if parameters $(\bm\gamma^+, \bm\eta^+, \bm\gamma^-,\bm\eta^-, \bm{Z}\bm{Z}^\top)$ can be uniquely determined from $\bm\Theta^+$ and $\bm\Theta^-$, where $\bm{Z}$ can be uniquely determined up to column permutation.

\begin{theorem}
\label{thm:ident}
SBBM is identifiable when $K=2$, if $\eta_1^+=0$, $\eta_1^-=0$ and there does not exist $S_1 \subseteq C_1$ and $S_2 \subseteq C_2$ such that $\eta_i^++\eta_j^+=0$ or $\eta_i^-+\eta_j^-=0$ for any $i \in S_1, j\in S_2$ and $i\in C_1\setminus S_1, j\in C_2\setminus S_2$.

SBBM is also identifiable when $K\ge 3$, if \\
\indent (i) there does not exist a community $C$ and a subset $S \subseteq C$ such that $\eta_i^++\eta_j^+=0$ or 
 $\eta_i^-+\eta_j^-=0$ for any $i\in S$ and $j\in C\setminus S$; and \\
\indent (ii) there does not exist two communities $C$ and $C'$ such that $\eta_i^++\eta_j^+=0$ or 
 $\eta_i^-+\eta_j^-=0$ for any $i\in C$ and $j\in C'$.
\end{theorem}

Intuitively, SBBM is unidentifiable if the difference between \(\bm\beta^+\) and \(\bm\gamma^+\), or between \(\bm\beta^-\) and \(\bm\gamma^-\) is not substantial, especially when their effects intertwine with the community membership. Theorem \ref{thm:ident} establishes the identifiability of SBBM by assuring that there are not too many node pairs with $\eta_i^++\eta_j^+ = 0$ or $\eta_i^-+\eta_j^- = 0$ so that the connection probability within the same community and across different communities can be well discriminated \citep{latouche2011overlapping, zhang2022identifiability}. The case with \(K=2\) requires separate treatment, as the underlying network distribution remains invariant when all \(\eta^+\)'s or \(\eta^-\)'s in one community increase but decrease in the other. Proof of Theorem \(\ref{thm:ident}\) involves a detailed graph-theoretic analysis of the intertwined relationship between the parameters and community membership, which may be of independent interest.

\subsection{Computing algorithm}\label{sec:computing}

The negative log-likelihood of $\bm A^+ = (a^+_{ij})_{i, j=1}^n$ and $ \bm A^- = (a^-_{ij})_{i, j=1}^n$ can be written as
$$
\mathcal{L}_n(\bm\Theta^+, \bm \Theta^-)=-\sum_{ 1\leq i, j\leq n}\Big(a^+_{ij}\theta^+_{ij}+a^-_{ij}\theta^-_{ij}-\log\big(1+e^{\theta^+_{ij}}+e^{\theta^-_{ij}}\big)\Big), 
$$
where $a_{ij}^{+}=\mathbb{I}(a_{ij}=1)$ and $a_{ij}^{-}=\mathbb{I}(a_{ij}=-1)$. Then we have 
\begin{align}\label{def:obj_norm}
	(\widehat{\bm{\Theta}}^+, \widehat{\bm{\Theta}}^-) = \operatorname*{arg\,min}_{\bm{\Theta}^+, \bm{\Theta}^-} \left\{ \mathcal{L}_n(\bm\Theta^+, \bm\Theta^-) + \lambda_n^+\|\bm\Theta^+\|_* + \lambda_n^-\|\bm\Theta^-\|_* \right\},
\end{align}
where $\|\cdot\|_*$ denotes the nuclear norm, which is used to promote the low-rank structure in $\bm{\Theta}^+$ and $\bm{\Theta}^-$, and $\lambda_n^+$ and $\lambda_n^-$ are two tuning parameters. The objective function in \eqref{def:obj_norm} is convex, and it can be efficiently optimized via the accelerated proximal gradient algorithm \citep{beck2009fast}. The algorithm iterates between a gradient descent step on the smooth loss $\mathcal{L}_n$ and a proximal step for the nuclear norm, which is computed via the singular value thresholding operator \citep{cai2010singular}, $\mathcal{D}_{\tau}(\bm M) = \sum_{k} \max\{\sigma_k - \tau, 0\} \bm u_k \bm v_k^\top$ for any matrix $\bm M$ with singular value decomposition $\bm M = \sum_{k} \sigma_k \bm u_k \bm v_k^\top$. Once $\widehat{\bm \Theta}^+$ and $\widehat{\bm\Theta}^-$ are obtained, we then proceed to estimate $\sigma$.

\begin{proposition} \label{prop:line_K}
Let $\bm \Theta_{\Delta} = \bm \Theta^+-\bm \Theta^-$ and $\bm J\bm \Theta_{\Delta} \bm J = \bm Y\bm \Sigma\bm Y^\top$ be the eigen decomposition with $\bm Y\in \bbR^{n\times r}$ and $r=\text{rank}\ (\bm J\bm \Theta_{\Delta} \bm J)$. Then there are $K$ lines in $\bbR^r$, denoted as $\{l_k:\bm w_k+t\bm v_k, t\in \bbR, \bm w_k, \bm v_k\in \bbR^r, k=1, \dots, K \}$, such that each row of $\bm Y$ lies exactly on one of these $K$ lines. Furthermore, for any $i, j\in[n]$, if $\sigma(i)=\sigma(j)\in[K]$, then $\bm Y_{i\cdot}$ and $\bm Y_{j\cdot}$ live on the same line.  
\end{proposition}

Proposition \ref{prop:line_K} implies that the community structure in $\sigma$ can be estimated from the spectral decomposition of $\bm J\widehat{\bm \Theta}_{\Delta}\bm J$, where $\widehat{\bm \Theta}_{\Delta}=\widehat{\bm \Theta}^+ -  \widehat{\bm \Theta}^-$. Denote the eigenvectors of $\bm J\widehat{\bm \Theta}_{\Delta}\bm J$ as $\widehat{\bm Y}=(\widehat{\bm Y}_1,\ldots, \widehat{\bm Y}_n)^\top$, then we can recover $\sigma$
via a $K$-lines clustering of $\widehat{\bm Y}$. Formally, given $\{\widehat{\bm Y}_1,\ldots, \widehat{\bm Y}_n\}$ and a set of lines $\bm L = \{\ell_1, \dots, \ell_K\}$, we define
\begin{align*}
    \Phi(\widehat{\bm Y}, \bm L, \sigma) = \frac{1}{n} \sum_{i=1}^n\|\widehat{\bm Y}_i-P_{\ell_{\sigma(i)}}\widehat{\bm Y}_i\|_2^2, 
\end{align*}
where $P_{l}$ indicates the projection onto the $l$-th line. Let $\sigma_{\bm L}(i) = \arg\min_{k\in[K]} \|\widehat{\bm Y}_i-P_{l_k} \widehat{\bm Y}_i\|_2$, then $\Phi(\bm Y, \bm L, \sigma_{\bm L}) = \min_{\sigma}\Phi(\bm Y, \bm L, \sigma)$. Therefore, the $K$-line clustering aims to find $\widehat{\bm L} = \arg\min_{\bm L} \Phi(\widehat{\bm Y}, \bm L, \sigma_{\bm L})$, which becomes a projective clustering problem \citep{tukan2022new}. The general projective clustering problem is NP-hard, but various techniques have been developed to obtain its $(1+\epsilon)$-approximated solutions \citep{feldman2006coresets, ding2012linear, feldman2020turning}. In particular, we adopt the method in \cite{ding2012linear}, which guarantees an objective value no greater than $(1+\epsilon)$ times the optimal value over the entire search space. Denote the solution as $\widehat{\boldsymbol{L}} = \{ \widehat{\ell}_1, \dots, \widehat{\ell}_K \}$ such that 
\[
\Phi(\widehat{\boldsymbol{Y}}, \widehat{\boldsymbol{L}}, \sigma_{\widehat{\boldsymbol{L}}})
\le (1+\epsilon)\, \min_{\boldsymbol{L}} \Phi(\widehat{\boldsymbol{Y}}, \boldsymbol{L}, \sigma_{\boldsymbol{L}}),
\]
and then the community membership is estimated as $\widehat{\sigma} = \sigma_{\widehat{\boldsymbol{L}}}$.

\section{Numerical experiments}

In this section, we examine the numerical performance of the proposed SBBM model, and compare it against three popular competitors in the literature, including the signed Laplacian-based method \citep{kunegis2010spectral}, the SPONGE algorithm based on signed SBM \citep{cucuringu2021regularized}, and the joint estimation method based on the joint inner product model \citep{tang2024population}, denoted as SLP, SPONGE and JIM, respectively. All methods are evaluated in terms of the community detection error. 

\subsection{Simulated examples}


\textbf{Example 1.} Following the data generating scheme in \cite{tang2024population}, we first generate $\bm{X} \in \mathbb{R}^{n \times 2}$ with rows independently set to $(1,0)$ or $(0,1)$ with equal probability $0.5$, and then set $\bm{X}^* = (\bm{I}_n - \tfrac{1}{n}\mathbf{1}\mathbf{1}^\top)\bm{X}$. For each node $i \in [n]$, we generate $\alpha_i^* \sim \mathrm{Unif}(1,3)$ and let $v_i^* = (1,-1)^\top \bm{x}_i^*/\sqrt{2}$, where $\bm{x}_i^*$ denotes the $i$-th row of $\bm{X}^*$. Edge probabilities are then specified by $\Pr(|A_{ij}|=1)=\sigma( \mu^* + \alpha_i^* + \alpha_j^* - 2\bar{\alpha}^* + \bm{x}_i^{*\top}\bm{x}_j^*)$ and $\Pr(A_{ij}=1 \mid |A_{ij}|=1)=\sigma( v_i^* v_j^*)$, where $\bar{\alpha}^* = n^{-1}\sum_{i=1}^n \alpha_i^*$ and $\sigma(x)=(1+e^{-x})^{-1}$. Varying $\mu^* \in \{-4.0,-3.5,-3.0\}$ produces strong-balanced signed networks with different sparsity.



We consider $n \in \{500,1000\}$. The averaged community detection errors, along with their standard errors, over 50 independent replications are reported in Tables \ref{table:1}

\begin{table}[!htb]
\centering
\small
\caption{Averaged community detection errors and their standard errors of all methods over 50 independent replications in Example 1. The best performer in each scenario is boldfaced.}
\label{table:1}
\begin{tabularx}{0.9\textwidth}{c c|>{\centering\arraybackslash}X >{\centering\arraybackslash}X >{\centering\arraybackslash}X >{\centering\arraybackslash}X}
\toprule
$n$ & $\mu$ & SBBM & SLP & SPONGE & JIM \\
\midrule
\multirow{3}{*}{500} & -4.0 & \textbf{0.1392(0.0076)} & 0.4481(0.0066) & 0.4370(0.0080) & 0.2618(0.0204) \\
\cmidrule(l){2-6}
 & -3.5 & \textbf{0.0909(0.0025)} & 0.3067(0.0095) & 0.2992(0.0160) & 0.1536(0.0233) \\
\cmidrule(l){2-6}
 & -3.0 & \textbf{0.0632(0.0022)} & 0.1712(0.0047) & 0.1239(0.0033) & 0.0692(0.0153) \\
\midrule
\multirow{3}{*}{1000} & -4.0 & \textbf{0.0860(0.0024)} & 0.2357(0.0051) & 0.2187(0.0137) & 0.1332(0.0201) \\
\cmidrule(l){2-6}
 & -3.5 & 0.0578(0.0019) & 0.1138(0.0019) & 0.0826(0.0017) & \textbf{0.0481(0.0140)} \\
\cmidrule(l){2-6}
 & -3.0 & 0.0414(0.0011) & 0.0567(0.0013) & 0.0369(0.0010) & \textbf{0.0121(0.0032)} \\
\bottomrule
\end{tabularx}
\end{table}

\bibliography{ref_signed}

\end{document}